# 1M PARAMETERS ARE ENOUGH? A LIGHTWEIGHT CNN-BASED MODEL FOR MEDICAL IMAGE SEGMENTATION

A PREPRINT

**Binh-Duong Dinh, Thanh-Thu Nguyen, Thi-Thao Tran*, Van-Truong Pham**
Department of Automation Engineering
School of Electrical and Electronic Engineering
Hanoi University of Science and Technology
duong.db190043@sis.hust.edu.vn, thu.nt192096@sis.hust.edu.vn, thao.tranthi@hust.edu.vn,
truong.phamvan@hust.edu.vn

## ABSTRACT

Convolutional neural networks (CNNs) and Transformer-based models are being widely applied in medical image segmentation thanks to their ability to extract high-level features and capture important aspects of the image. However, there's often a trade-off between the need for high accuracy and the desire for low computational cost. A model with higher parameters can theoretically achieve better performance but also result in more computational complexity and higher memory usage, and thus is not practical to implement. In this paper, we look for a lightweight U-Net-based model which can remain the same or even achieve better performance, namely U-Lite. We design U-Lite based on the principle of Depthwise Separable Convolution so that the model can both leverage the strength of CNNs and reduce a remarkable number of computing parameters. Specifically, we propose Axial Depthwise Convolutions with kernels $\mathbf{7 \times 7}$ in both the encoder and decoder to enlarge the model's receptive field. To further improve the performance, we use several Axial Dilated Depthwise Convolutions with filters $\mathbf{3 \times 3}$ for the bottleneck as one of our branches. Overall, U-Lite contains only 878K parameters, 35 times less than the traditional U-Net, and much more times less than other modern Transformer-based models. The proposed model cuts down a large amount of computational complexity while attaining an impressive performance on medical segmentation tasks compared to other state-of-the-art architectures. The code will be available at: https://github.com/duong-db/U-Lite.

## 1 Introduction

Artificial Intelligence (AI) has been recently deployed in many hospitals for practical use in developed countries. According to Health Equity Magazine (2018), AI helps store and access a large amount of information effectively. Currently, the amount of medical information has doubled every three years. It is estimated that if a doctor wants to keep up to date with all medical news, he has to read 29 hours a day, which is impossible. On the other hand, a deep learning model can also assist in diagnosing medical images. The IBM Watson healthcare system (USA) shows that 90% of the recommendations made by their AI system are consistent with the suggestions of medical experts, but taking only 40 seconds to complete all processes. Realizing the importance of AI in the medical field, many study efforts have been made to improve the performance of deep learning models.

Computer Vision is one of the most prominent applications of AI in the medical field. Besides diagnosing diseases based on clinical symptoms, modern medicine also diagnoses thediseases based on subclinical symptoms from images obtained from medical devices. Therefore, deep learning models are developed for segmenting tumors and cells or abnormal areas, thereby initially supporting doctors in the process of disease identification and diagnosis as well as the severity of the disease. In 2015, Ronneberger et al. introduced U-Net [1] as a highly effective model for medical image segmentation. After the success of U-Net, many follow-ups works did different research to optimize U-Net and offered numerous variants with higher performance such as Unet++ [2], ResUnet++ [3], Double Unet [4], Attention Unet [5], to name just a few. By and large, they all are deep learning models developed based on CNNs. There is no denying that the appearance of CNNs has created a great revolution in computer vision fields.

Recently, Vision Transformer [6] and MLP-like architectures (MLPs) have been widely used and became a new de-facto standard in Computer Vision. Vision Transformer considers each patch of an image as a token and feeds them through the Multi-head Self-attention mechanism as they successfully did with sentences in Natural Language Processing (NLP). In the medical segmentation task, TransUnet [7] can be considered one of the high-performing models in accuracy and efficiency. Following the success of TransUnet, transformer-based models continued to be developed. Pyramid Vision Transformer (PVT) [8] is used as the backbone of many high-performance models such as MSMA-Net [9], Polyp-PVT [10]. Meanwhile, MLP-like architectures are also very research focused. MLPs leverage the advantage of conventional MLP to encode the features along each of their dimensions. AxialAtt-MLP-Mixer [11] gives very good performance on many medical image datasets by applying axial attention to replace the token mixing in MLP-Mixer [12]. Different from CNNs, the models based on transformers or MLPs mainly concentrate on the global receptive field of the image, thus costing much in computational complexity and being overly heavy for the training process.

To successfully implement in practice, a machine learning model first needs to achieve high accuracy and secondly, it should be fast and compact enough to integrate into mobile medical devices. Nevertheless, there's often a trade-off between the need for high accuracy and the desire for low computational cost. These above studies can theoretically achieve impressive performance, however, a large number of them may give heavy operation and slow calculation speed due to the massive number of parameters. To solve this problem, some attempts for a lightweight architecture can be mentioned as Mobile-Unet [13], DSCA-Net [14], and MedT [15]. In this paper, we rethink an efficient lightweight architecture for the medical segmentation task to further explore a high-performing model which can effectively address this issue. In short, the main contributions of this paper are three-fold:

1. Propose the usage of Axial Depthwise Convolution module based on the concept of Depthwise Separable Convolution. This module helps the model solve every complex architecture problem: enlarging the model's receptive field while reducing the heavy computation burden.

2. Propose U-Lite, a lightweight and simple CNN-based architecture. As far as we know, U-Lite is one of the few models that surpasses a recent highly efficient compact network UneXt [16] in terms of both performance and the number of parameters.

3. We have successfully implemented the model on medical segmentation datasets and achieved considerable results.

## 2 Related work

**U-Net.** Introduced in 2015 [1], this is an effective deep-learning model based on CNNs for biomedical image segmentation. U-Net follows U-shape architecture using a typical encoder-decoder structure. The encoder of U-Net extracts high-level features through a combination of convolution layers, activation functions, and normalized layers consecutively. Besides, thanks to the max-pooling layers, the encoded information extracted from the input image is multi-scale which is suitable for a variety of objects with different shapes and sizes such as tumors, cells, ... in the medical image segmentation task. After obtaining multi-scale feature representations from the encoder, the decoder takes several upsampling steps to restore the original size and then compares the output predicted mask with the ground truth. To avoid local information loss problem affected by max-pooling layers, skip connections are proposed. This helps combining spatial representations between features of the encoder and decoder, thereby increasing the performance and solving the problem of vanishing gradients.

**Depthwise Separable Convolution.** Depthwise Separable Convolution consists of two convolutions namely depthwise convolution and pointwise convolution. Depthwise convolution does not change the depth of the feature map, each kernel of this convolution is applied for each channel individually instead of all channels like normal convolution. After that, a pointwise convolution is applied with a $1 \times 1$ kernel size to change the depth of the feature map. Depthwise Separable Convolution gives the same performance as traditional convolution, but it uses fewer parameters, thereby reducing computational complexity and making the model more compact. Therefore, they are used quite a lot in modern deep learning models, such as Xception [17] and MobileNets [18].

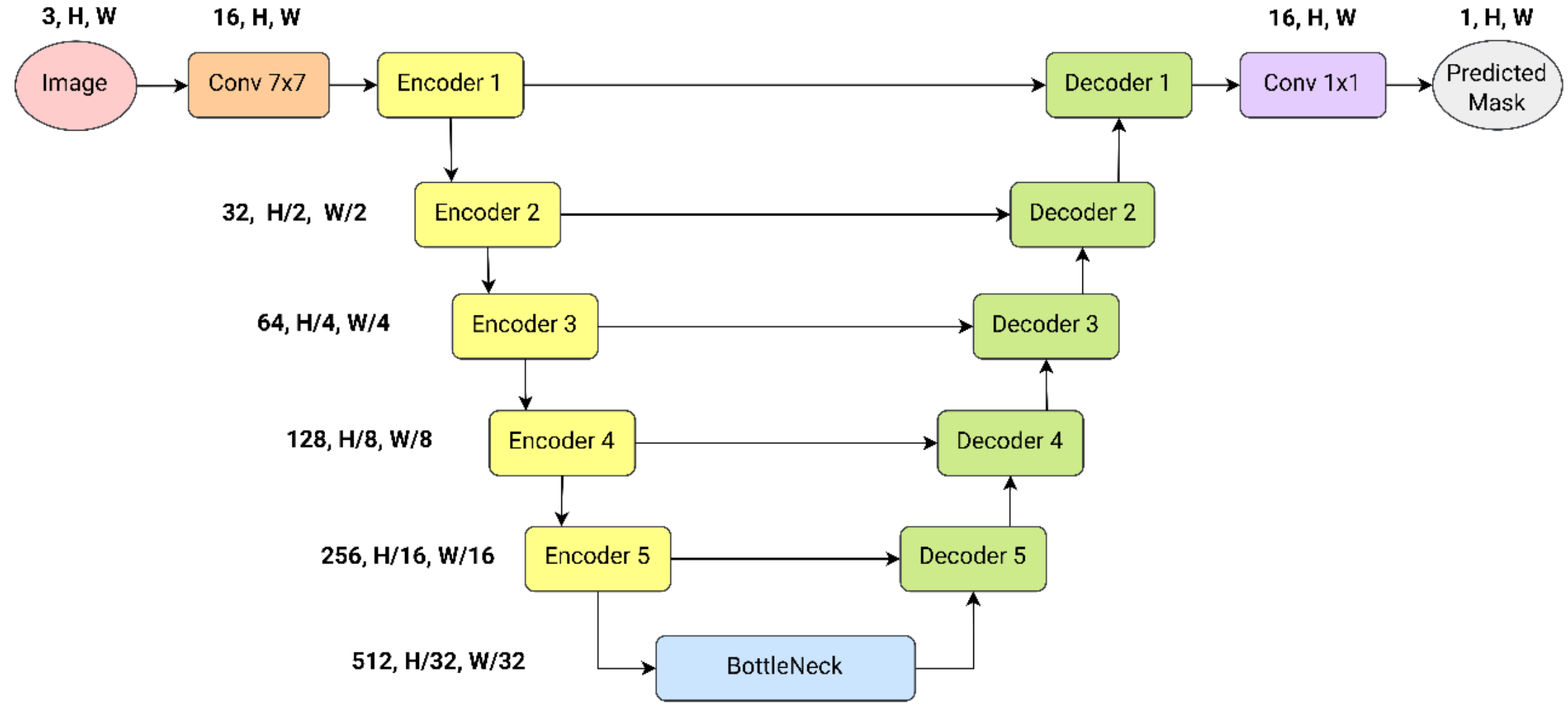

Figure 1: The proposed U-Lite architecture.

## 3 Proposed Model

An overview of our proposed model U-Lite is illustrated in Fig. 1. We follow the symmetric encoder-decoder architecture of U-Net and design U-Lite in an efficient way so that the model can leverage the strength of CNNs while maintaining the number of computational parameters as small as possible. To this end, an Axial Depthwise Convolution module is thoughtfully proposed, shown in Fig. 2. Describing the operation of U-Lite, an input image at shape (3*, H, W*) is fed to the network through 3 stages: encoder stage, bottleneck stage, and decoder stage. U-Lite follows the hierarchical architecture, where the encoder extracts six different level features in shape $\left(C_i, \frac{H}{2^i}, \frac{W}{2^i}\right)$ where $i \in \{0, 1, 2, ..., 5\}$. The bottleneck and decoder take part in processing these features as well as upscaling them to the original shape to obtain the segmented mask. We also use skip connections between the encoder and decoder. It is worth noting that although the design of U-Lite is simple, the model still performs well on the segmentation task thanks to the contribution of Axial Depthwise Convolution module.

### 3.1 Axial Depthwise Convolution module

The success of Vision Transformer [6] promotes various works on researching and improving this special structure. Swin Transformer [19] reduces the computational complexity of Transformer by limiting self-attention computation to non-overlapping local windows of size 7 × 7. ConvNext [20] realizes this modification and adopts convolutions with kernel size 7 × 7 to CNN architecture, bringing ResNet up to 86.4% top-1 accuracy on ImageNet. Meanwhile, a recent new paradigm, Vision Permutator [21] makes use of linear projections to separately encode the feature representations along the height and width dimensions. This variant of MLP-like architecture is supposed to readily attain promising results in Computer Vision. Our exploration is motivated by a natural question: *What would happen if we replace the cruciform receptive field of Vision Permutator with a local receptive field version, in the same manner as Swin Transformer did with Vision Transformer?*

To give a simple answer, we propose Axial Depthwise Convolution module, as a combination of Vision Permutator's and ConvNext's designs (Fig. 2). The mathematical formulation of this operator is expressed as follows:

$$x' = x + DW_{1\times n}(x) + DW_{n\times 1}(x) \tag{1}$$

$$y = GELU\left(PW_{C_1 \to C_2}\left(BN(x')\right)\right) \tag{2}$$

where: $x$ is input feature, $y$ is output feature; $DW$, $PW$ and $BN$ stand for Depthwise Convolution, Pointwise Convolution, and Batch Normalization respectively, $1 \times n$ and $n \times 1$ are the kernel sizes of the convolutions; $C_1$ and $C_2$ represent the number of input and output channels of the feature map. In our experiment, n = 7. To achieve a minimal and flexible design, we use a unique Pointwise Convolution without adding residual connection, allowing to change the number of input channels adaptively.

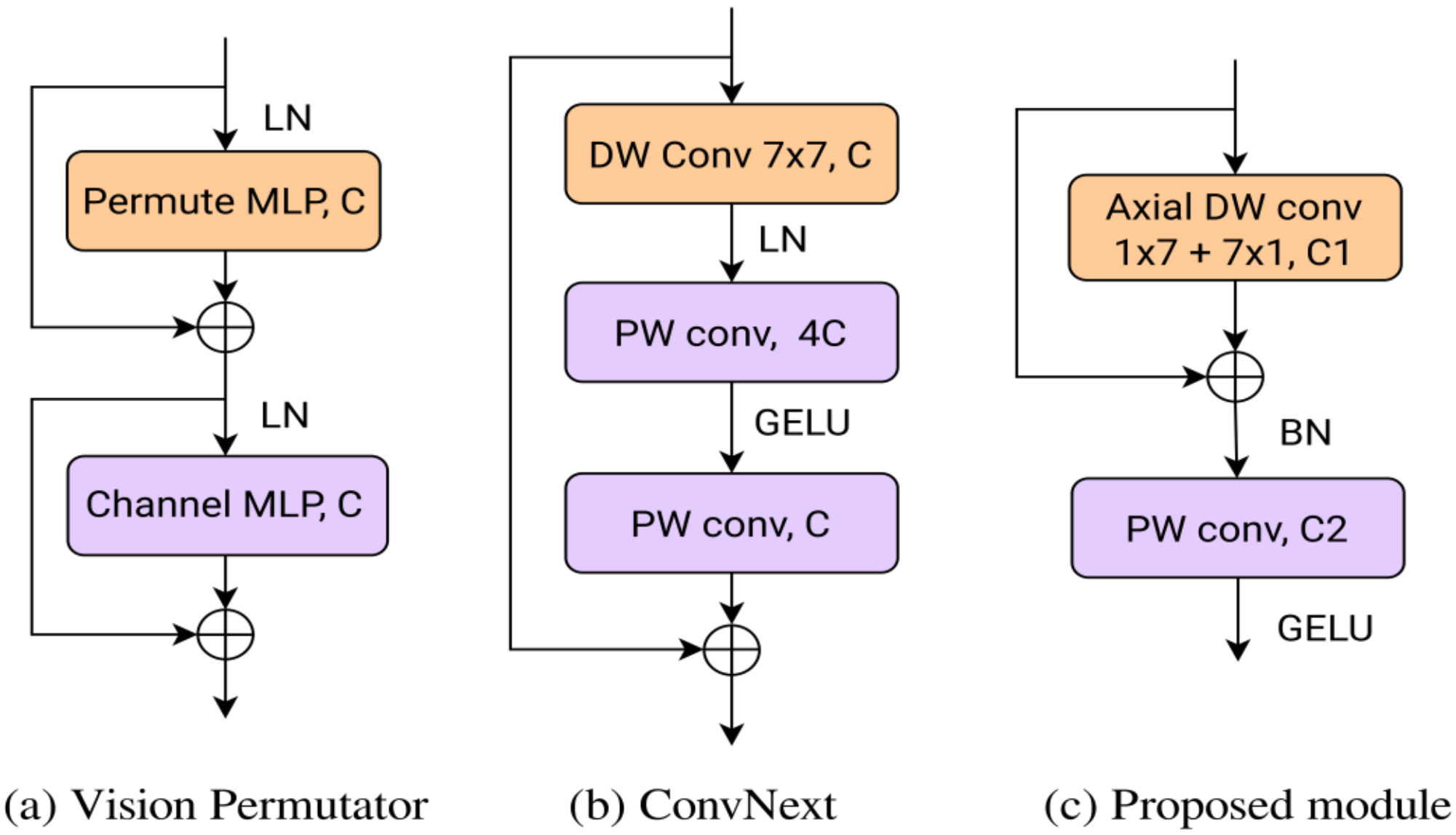

Figure 2: Architectures of (a) Vision Permutator, (b) ConvNext, and (c) proposed Axial DW Convolution module. The proposed module is inspired from Vision Permutator's and ConvNext's designs.

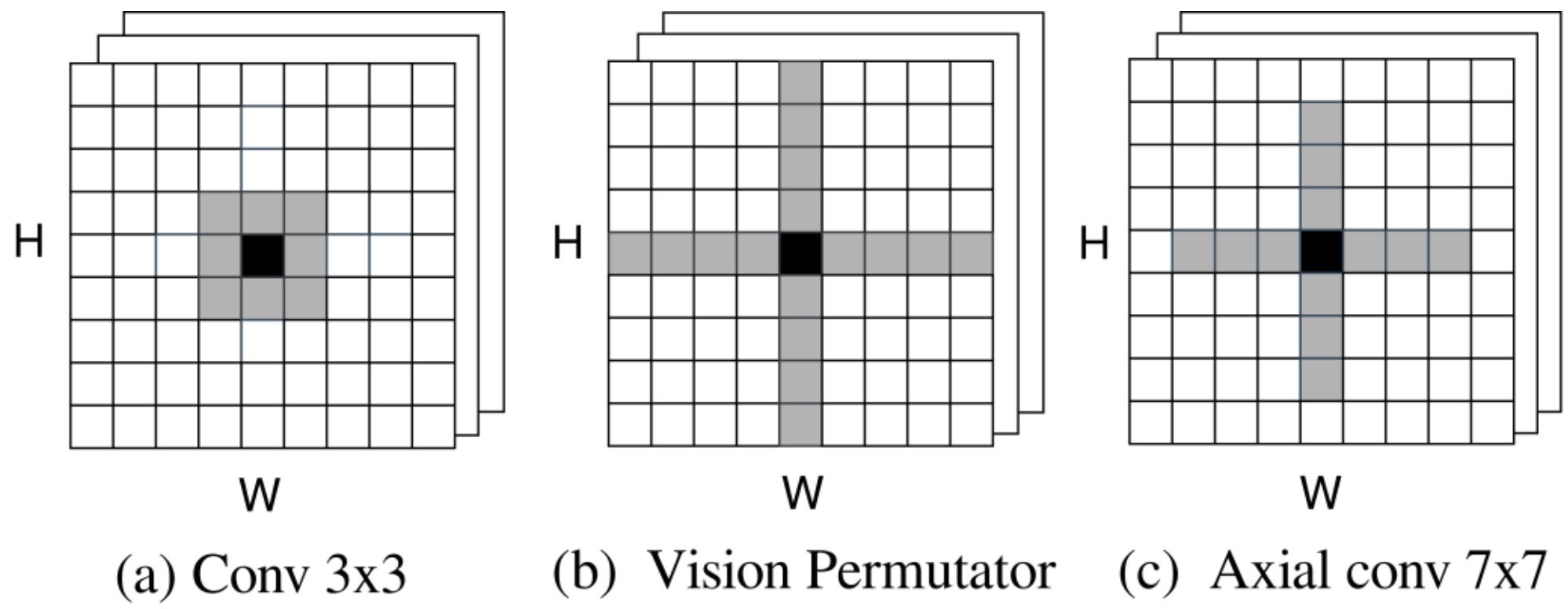

Figure 3: The receptive field comparison between Convolution $3 \times 3$, Vision Permutator, and Axial convolution $7 \times 7$. Axial convolution 7x7 offers a large receptive field compared with Convolution $3 \times 3$ while using less computational parameters than Vision Permutator.

### 3.2 Encoder Block and Decoder Block

The design principles of the encoder and decoder blocks are established as follows:

- Follow the Depthwise Separable Convolution architecture. This is an important key to successfully building a lightweight model from scratch. Depthwise Separable Convolution gives the same performance as traditional convolution while using fewer parameters, therefore reducing computational complexity and making the model more compact.

- Limit the use of unnecessary operators. Simply using normal MaxPooling and UpSampling layers. There is no need for a high-parameter-consuming operator such as Transposed Convolution. A Pointwise Convolution operator can play two roles simultaneously: encoding features along the depth of the feature map while flexibly changing the number of input channels.
- Each encoder or decoder block adopts one Batch Normalization layer and ends with a GELU activation function. We have made a performance comparison between Batch Normalization and Layer Normalization but there is not much difference. GELU is applied since it is proved to give an improvement in accuracy when using GELU compared to ReLU, and ELU.

The encoder and decoder structures of U-Lite are shown in Fig. 4.

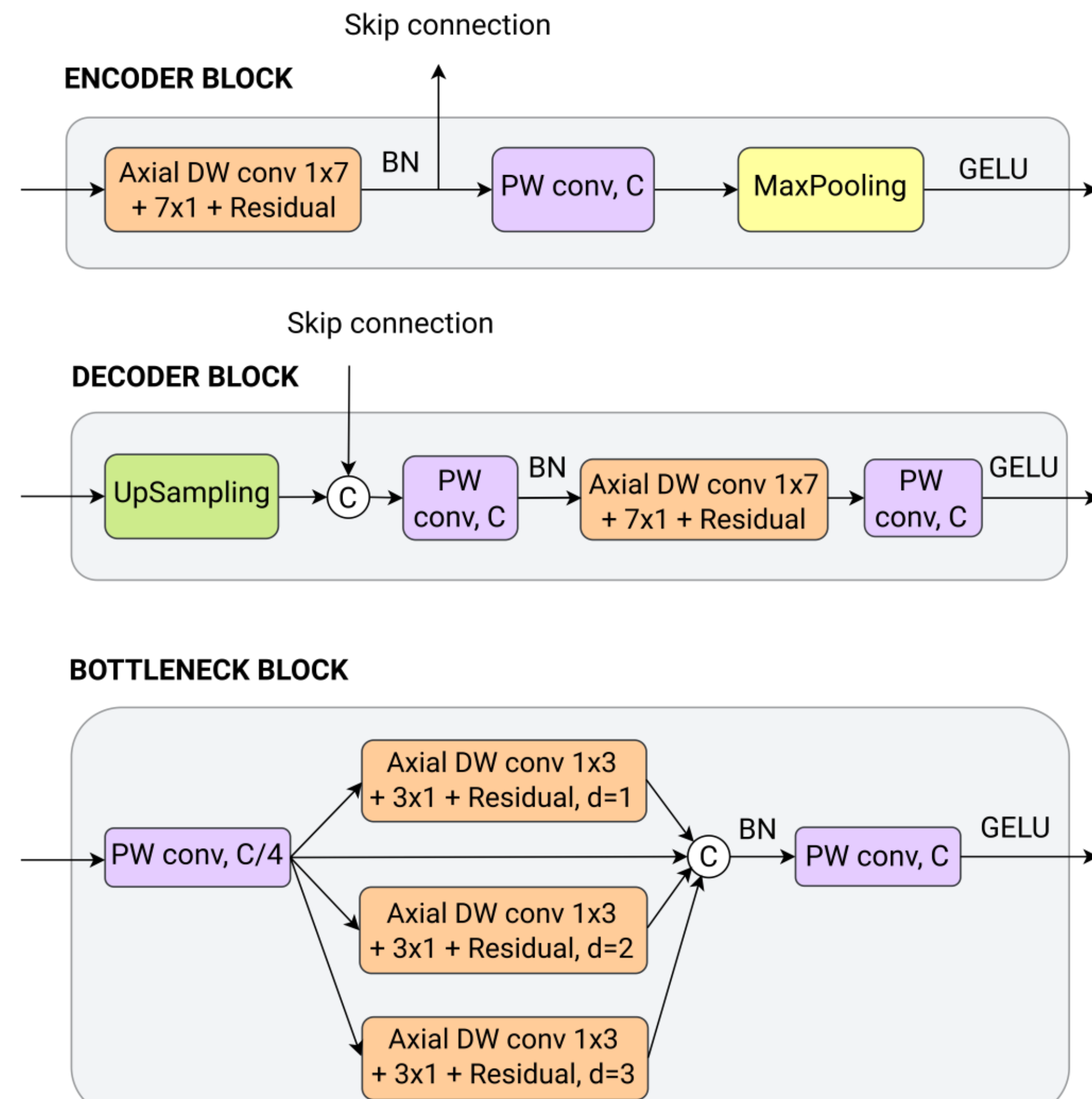

Figure 4: Encoder, decoder, and bottleneck blocks. Designing based on Depthwise Separable Convolution concept. Each block adopts one Batch Normalization layer and ends with a GELU activation function.

### 3.3 Bottleneck Block

To further boost the performance of U-Lite, we apply Axial Dilatied Depthwise Convolutions with kernel size $n = 3$ to the bottleneck block (Fig. 4). The applied dilation rates are $d = 1,2,3$. We use axial dilated convolutions with a kernel of size 3 for two reasons: 1) a kernel of size 3 is more suitable with the spatial shape of bottom-layer features, where the height and width of such features are reduced multiple times, 2) it gives a better performance when using dilated convolutions with different rates to capture multi-spatial representations of high-level features at the bottom stage. To further reduce the number of learnable parameters, a pointwise convolution layer is adopted at the beginning of the bottleneck block. This helps scaling down the channel dimension of last layer features before feeding them to the Axial Dilated Depthwise Convolution mechanism.

# 4 Experiments

## 4.1 Implementation detail

**Dataset.** We implement the training on three datasets to verify the segmentation performance of the proposed model. First, the ISIC2018 dataset consists of 2594 images of abnormal areas of the skin. We divide this dataset into 2334 images for training and 260 images for testing. The Data Science Bowl (DSB) 2018 dataset consists of 617 nuclear snapshots of cells, which is divided into 80% for training and 20% for testing. The final dataset we used in this study is the GlaS (Gland Segmentation) dataset. The GlaS dataset consists of 165 colorectal adenocarcinoma images, of which 85 are used for training and 80 for testing. In all the experiments, the images are resized to 256 × 256.

**Training strategy.** We implement U-Lite on PyTorch framework using 16GB NVIDIA Tesla T4 GPU. To avoid over-fitting during the training phase with 200 epochs without pretrained, data is augmented by a number of techniques including random rotation, horizontal flip, and vertical flip. We employ Adam optimization [24] algorithm at the initial learning rate $lr = 1e-3$ and use Dice loss function. The Dice Loss evaluates the overlap between the prediction and the ground truth of an image. The mathematical formulation of Dice Loss is represented as follows:

$$L_{Dice}(G,P) = 1 - \frac{2\sum_{i=1}^{N} G_i P_i}{\sum_{i=1}^{N}(G_i + P_i)} \tag{3}$$

where $N$ is the total pixels of the image, $G_i \in \{0,1\}$ denotes the ground truth label of the $i$-th pixel and $P_i \in \{0,1\}$ presents probability prediction score for $i$-th pixel.

**Evaluation Metric.** In order to evaluate the model performance, we use the two most popular metrics in the semantic segmentation task, namely Dice Similarity Coefficient (Dice) and Intersection over Union (IoU). Both Dice and IoU are used to evaluate the similarity between the ground truth and the predicted mask. The mathematical formulation of Dice and IoU are expressed as follows:

$$\text{Dice} = \frac{2\text{TP}}{2\text{TP} + \text{FP} + \text{FN} + \varepsilon} \tag{4}$$

$$\text{IoU} = \frac{\text{TP}}{\text{TP} + \text{FP} + \text{FN} + \varepsilon} \tag{5}$$

where TP, FP, FN are True Positives, False Positives and False Negatives, respectively, and $\epsilon$ is the smooth coefficient to avoid zero division. In our experiment, the smooth coefficient $\epsilon = 1e-5$.

## 4.2 Representative Results

Fig. 5, Fig. 6, and Fig. 7 show the qualitative segmentation of some representative images in test sets of various medical datasets including the ISIC 2018, DSB 2018, and GlaS respectively. To further show the performance of the proposed U-Lite model, we also present the segmentation results by the recent UneXt [16] model. It can be seen from the last two columns of each figure, the proposed U-Lite gives prediction masks closer to the ground truths compared with those by the UneXt, where the red circles mark the incorrect segmented areas that UneXt provided.

## 4.3 Comparative Results

To prove the outstanding performance of the proposed model, we compare the performance indicators of U-Lite with other proposed models in the medical segmentation task.

As shown in Table 1, for the ISIC 2018 dataset, U-Lite achieved 90.39% Dice and 83.63% IoU, with the Dice performs 2.02% higher than U-Net (a traditional CNNs model) and 0.53% higher than ConvUNeXt (a recent high-performing CNN-based model). It can be seen that the total parameters of the U-Lite model are very small, only 878K parameters, 35 times less than U-Net. The number of parameters of U-Lite is also smaller than that of some recent proposed lightweight models: ConvUNeXt with 3.5M parameters and UneXt with 1.5M parameters.

Table 1: QUANTITATIVE EVALUATION RESULTS ON ISIC2018 DATASET COMPARED TO PREVIOUSLY PROPOSED MODELS

| Model | Year | Params | Dice | IoU |
|---|---|---|---|---|
| U-Net [1] | 2015 | 31.2M | 88.60 | 81.58 |
| Attention Unet [5] | 2018 | 31.4M | 88.66 | 81.68 |
| Unet++ [2] | 2018 | 9.2M | 87.78 | 80.72 |
| ConvUNeXt [22] | 2022 | 3.5M | 89.81 | 83.19 |
| UneXt [16] | 2022 | 1.5M | 89.56 | 82.97 |
| U-Lite (Ours) | 2023 | **878K** | **90.39** | **83.63** |

Table 2 and Table 3 evaluate the performance of U-Lite on DSB 2018 and GlaS datasets, respectively, compared to previously proposed models. On DSB 2018, our proposed model achieves 91.37% Dice and 83.63% IoU. Meanwhile, on GlaS, these indices are 86.65% Dice and 77.84% IoU. These results are all higher than some of the models mentioned in the tables.

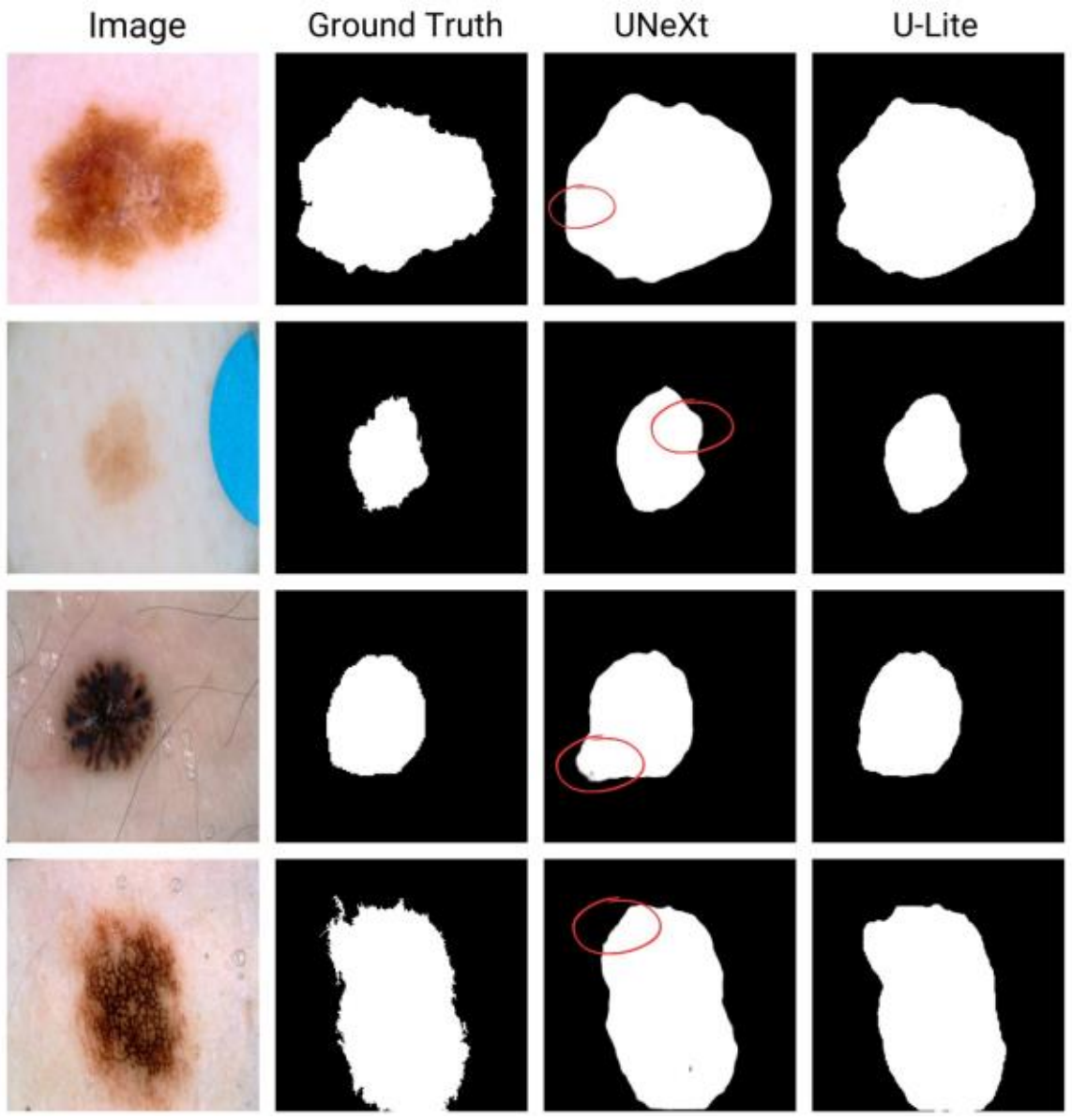

Figure 5: Qualitative comparisons between UNeXt and U-Lite on ISIC 2018 dataset

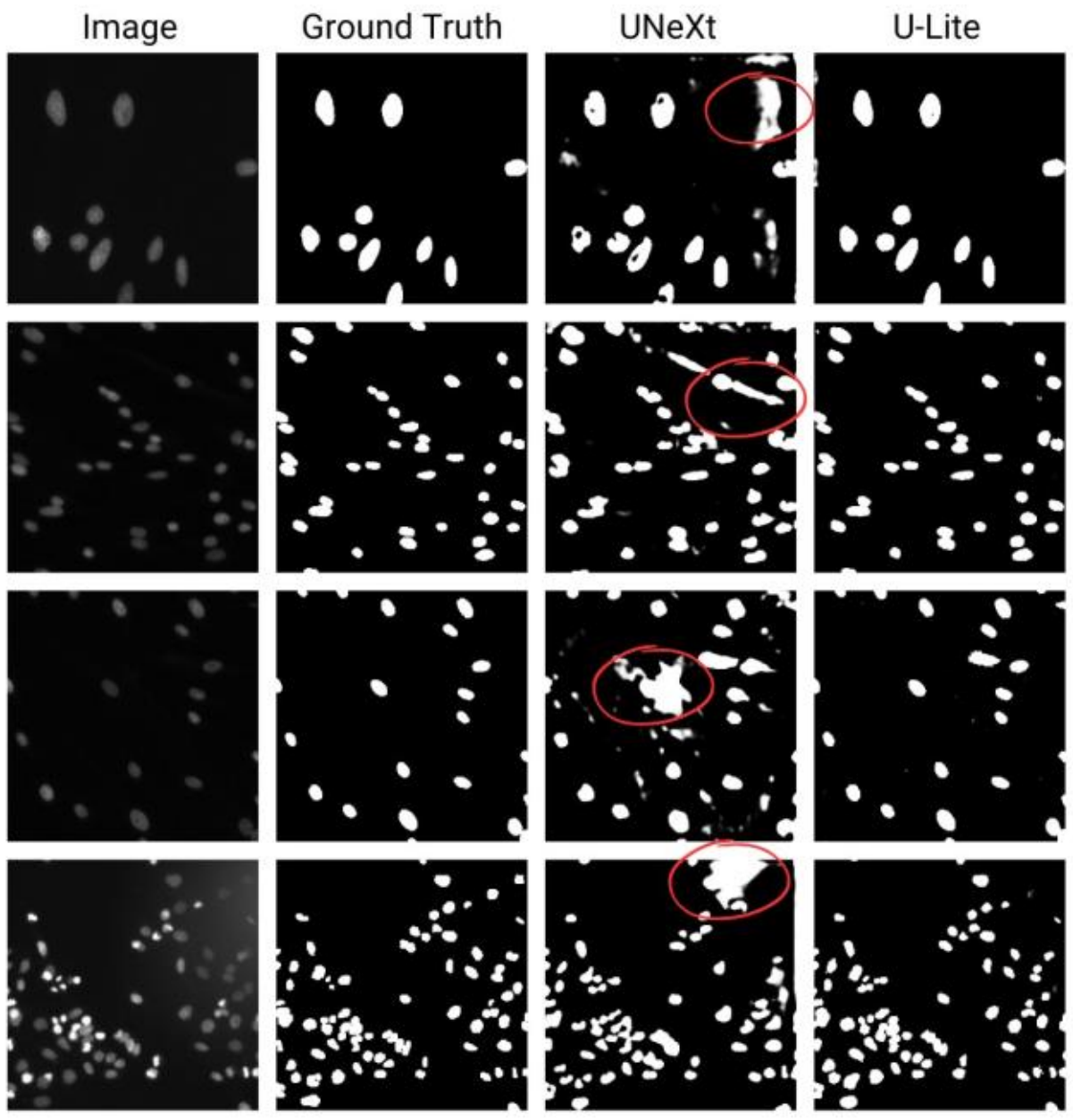

Figure 6: Qualitative comparisons between UNeXt and U-Lite on DSB 2018 dataset

## 5 Discussion

### 5.1 Analysis on the Axial Depthwise Convolution operator

To evaluate the performance of the proposed Axial Depthwise convolution in the proposed architecture, we replaced the Axial Depthwise Convolution operator, in Fig. 2, with the conventional Depthwise Convolution operator and experimented with various sizes of kernel to compare their performance. The final results are shown in Table 4. Surprisingly, even though the Depthwise Convolution operator provides a larger receptive field with the same size of kernel (i.e, same value of *n*), Axial Depthwise Convolution still performs better and again shows simplicity in parameter configuration and computational complexity.

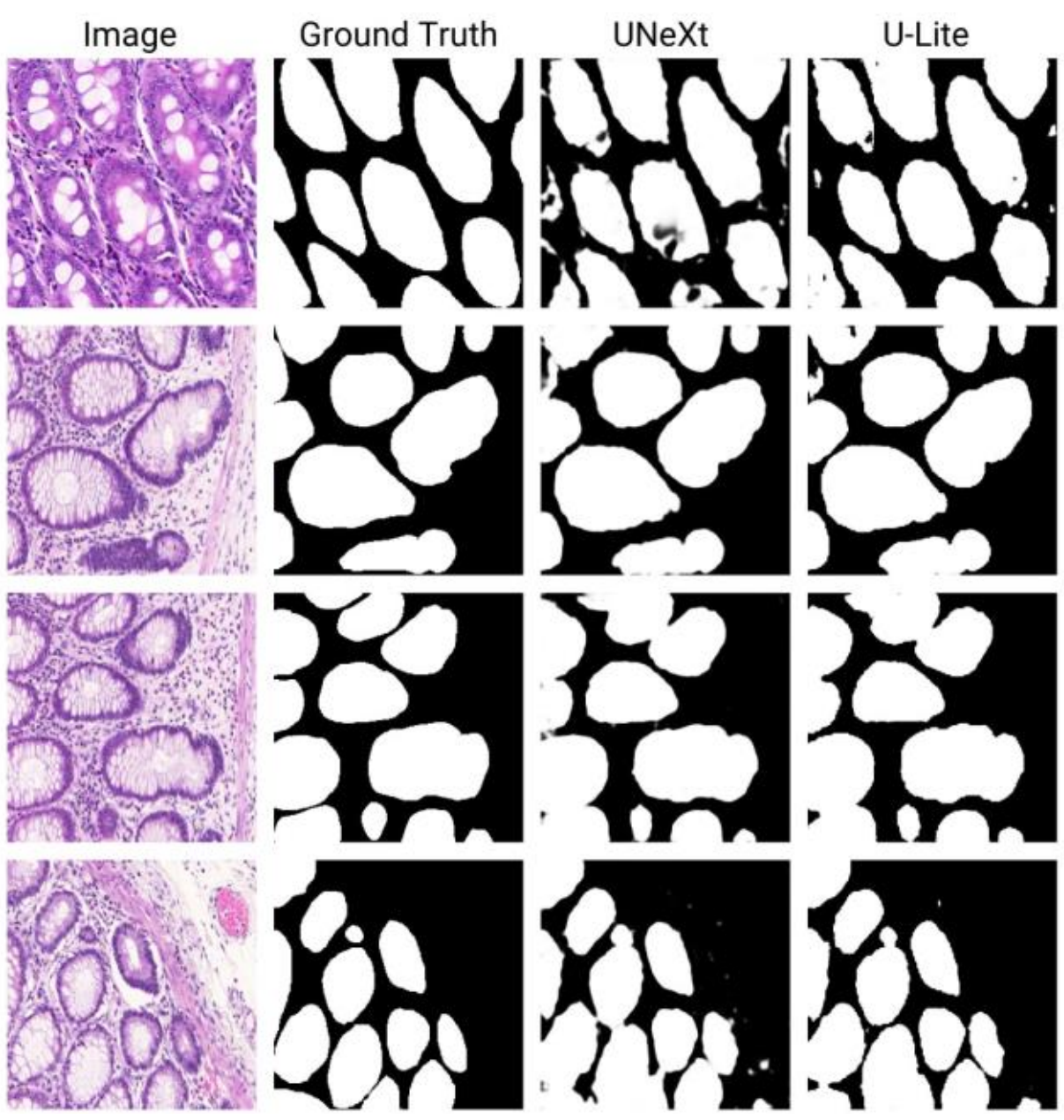

Figure 7: Qualitative comparisons between UNeXt and U-Lite on GlaS dataset

Table 2: QUANTITATIVE EVALUATION RESULTS ON DATA SCIENCE BOWL 2018 DATASET COMPARED TO PREVIOUSLY PROPOSED MODELS

| Model | Year | Params | Dice | IoU |
|---|---|---|---|---|
| U-Net [1] | 2015 | 7.9M | 88.93 | 82.57 |
| Unet++ [2] | 2018 | 9.2M | 88.96 | 82.77 |
| ConvUNeXt [22] | 2022 | 3.5M | 89.09 | 83.64 |
| UneXt [16] | 2022 | 1.5M | 88.13 | 82.15 |
| U-Lite (Ours) | 2023 | **878K** | **91.37** | **84.86** |

### 5.2 Effect of Axial Dilated Depthwise Convolutions to the bottleneck

Table 5 compares the performance of U-Lite before and after integrating the model with the Axial Dilated Depthwise Convolution (ADDC) mechanism. In the case of not using ADDC, we adopt the Axial Depthwise Convolution module with the kernel size $n = 7$ only to the bottleneck of U-Lite as an alternative. Overall, this model contains 817K parameters. It is observed that ADDC boosts the performance of U-Lite by 0.2 ÷ 0.5% in both Dice and IoU coefficients while not overly increasing the number of learnable parameters or the amount of computation.

Table 3: QUANTITATIVE EVALUATION RESULTS ON GLAS DATASET COMPARED TO PREVIOUSLY PROPOSED MODELS

| Model | Year | Params | Dice | IoU |
|---|---|---|---|---|
| U-Net [1] | 2015 | 7.9M | 77.78 | 65.34 |
| Unet++ [2] | 2018 | 9.2M | 78.03 | 65.55 |
| MedT [15] | 2022 | 1.6M | 81.02 | 69.61 |
| AxialAtt-MLP-Mixer [11] | 2022 | 29.0M | 84.99 | 73.97 |
| UneXt [16] | 2022 | 1.5M | 86.49 | 77.77 |
| U-Lite (Ours) | 2023 | **878K** | **86.85** | **77.84** |

Table 4: PERFORMANCE COMPARISON BETWEEN DEPTHWISE CONVOLUTION AND AXIAL DEPTHWISE CONVOLUTION WITH DIFFERENT SIZES OF KERNEL

| Dataset | Operator | Metric | n = 3 | n = 5 | n = 7 |
|---|---|---|---|---|---|
| ISIC 2018 | DW Conv | Dice | 89.71 | 89.91 | 89.89 |
| | | IoU | 82.92 | 83.19 | 83.48 |
| | Axial DW Conv | Dice | 90.04 | 90.31 | **90.39** |
| | | IoU | 83.35 | **83.68** | 83.63 |
| DSB 2018 | DW Conv | Dice | 90.95 | 91.01 | 91.27 |
| | | IoU | 84.30 | 84.36 | 84.74 |
| | Axial DW Conv | Dice | 91.21 | 91.34 | **91.37** |
| | | IoU | 84.60 | 84.77 | **84.86** |
| GlaS | DW Conv | Dice | 85.88 | 85.87 | 86.14 |
| | | IoU | 76.45 | 76.53 | 76.86 |
| | Axial DW Conv | Dice | 86.00 | 85.80 | **86.85** |
| | | IoU | 77.20 | 76.32 | **77.84** |

Table 5: EFFECT OF AXIAL DILATED DEPTHWISE CONVOLUTIONS TO THE BOTTLENECK

| Dataset | ADDC | Dice | IoU |
|---|---|---|---|
| ISIC 2018 | w/o | 89.96 | 83.21 |
| | w | **90.39** | **83.63** |
| DSB 2018 | w/o | 91.21 | 84.60 |
| | w | **91.37** | **84.86** |
| GlaS | w/o | 84.54 | 74.82 |
| | w | **85.65** | **77.84** |

## 6 Conclusion

In this paper, we proposed a lightweight CNN-based architecture for the medical segmentation task. The suggested model U-Lite overcomes the limitation of many complex models that need a high number of parameters and thus results in much memory size and low inference speed. To this end, we proposed the Axial Depthwise Convolution module as a combination of Vision Permutator and ConvNext blocks. We further boost the performance of U-Lite by integrating an Axial Dilated Depthwise Convolution mechanism to the bottleneck of the model. Our experiments on various medical datasets show that U-Lite has a smaller number of parameters and costs less in computational complexity while also performing promising results.